\documentstyle[aps,prl,preprint,floats,epsfig]{revtex}

\begin{document}

\preprint{\tighten\vbox{\hbox{\hfil CLNS 99/1603}
                        \hbox{\hfil CLEO 99-2}
}}

\title{Measurement of Charm Meson Lifetimes}  

% Your author list ***DOES NOT*** go here!
% is goes below where you are instructed to insert it...
\author{CLEO Collaboration}
\date{\today}

\maketitle
\tighten

\begin{abstract} 
% Insert abstract here.
We report measurements of the  $D^0$, $D^+$, and $D^+_s$ meson
lifetimes using 3.7~fb$^{-1}$ of $e^+e^-$ annihilation data collected
near the $\Upsilon(4S)$ resonance with the CLEO detector. The 
lifetimes of the  $D^0$, $D^+$, and $D^+_s$ mesons are measured to be
$408.5 \pm 4.1 ^{+3.5}_{-3.4}$~fs, $1033.6 \pm  22.1 ^{+~9.9}_{-12.7}$~fs,
and $486.3 \pm 15.0 ^{+4.9}_{-5.1}$~fs, respectively.
The precisions of the charm meson lifetimes reported here are comparable
to those of the best previous measurements, and the systematic
errors are very different.
In a single experiment we find that the ratio of the
$D^+_s$ and $D^0$ lifetimes differs from one by more than 4.5 standard
deviations.
\end{abstract}
\newpage

{
\renewcommand{\thefootnote}{\fnsymbol{footnote}}

% Insert author and address list here
\begin{center}
G.~Bonvicini,$^{1}$ D.~Cinabro,$^{1}$ R.~Greene,$^{1}$
L.~P.~Perera,$^{1}$ G.~J.~Zhou,$^{1}$
S.~Chan,$^{2}$ G.~Eigen,$^{2}$ E.~Lipeles,$^{2}$
M.~Schmidtler,$^{2}$ A.~Shapiro,$^{2}$
% ...
\end{center}
 
\small
\begin{center}
$^{1}${Wayne State University, Detroit, Michigan 48202}\\
$^{2}${California Institute of Technology, Pasadena, California 91125}\\
% ...
\end{center}

\setcounter{footnote}{0}
}
\newpage

% Insert body of the text here.
The systematics of charm hadron lifetimes have played a central role in 
understanding heavy quark decays~\cite{theory}.
In this Letter we report new measurements of the lifetimes of the $D^0$,
$D^+$, and $D^+_s$ mesons. 
These charm meson ground states differ in the identity of the light antiquark,
{\it i.e.}, the $D^+$, $D^0$, and $D^+_s$ mesons are $c\bar{d}$, $c\bar{u}$,
and $c\bar{s}$ states, respectively.
Although the weak decay of the charm quark is responsible for the decays
of all three charm mesons, differences in the lifetimes indicate that the 
identity of the light antiquark also influences the rates of decay.
The large ratio~\cite{pdg} of the $D^+$ and $D^0$ lifetimes
($\tau_{D^+}/\tau_{D^0} \sim 2.5$) arises primarily from destructive
interference between different quark diagrams that contribute significantly
only to decay of the $D^+$~\cite{theory}.
This interference, as well as a number of smaller effects, which can cause
the $D_s^+$ and $D^0$ lifetimes to differ, appear in a systematic expansion,
in inverse powers of the charm quark mass, of the QCD contributions to the
charm decay amplitudes~\cite{theory}.
The results described in this Letter indicate that the ratio of the $D^+_s$
and $D^0$ lifetimes differs significantly from one, providing a quantitative
challenge for the theoretical description of charm meson decays.
These data were obtained in an $e^+e^-$ colliding beam environment,
where the event topologies and backgrounds are very different from those
encountered in the high energy fixed target experiments~\cite{fixed-target}
that have recently provided the most precise measurements of charm hadron
lifetimes~\cite{pdg}.

The results described in this Letter are based on an integrated luminosity
of 3.7 fb$^{-1}$ of $e^+e^-$ annihilation data recorded with the CLEO~II.V
detector near the $\Upsilon(4S)$ resonance at the Cornell Electron Storage
Ring (CESR). 
This luminosity corresponds to approximately 4.4 million recorded 
$e^+ e^- \to c\bar{c}$ events. 
The CLEO~II detector has been described elsewhere~\cite{cleoii}.  
The major component of the CLEO~II.V upgrade completed in November 1995 is
the SVX, the first multi-layer silicon vertex detector operating near the
$\Upsilon(4S)$ energy~\cite{svx}. 
The SVX consists of three concentric layers of 300~$\mu $m thick,
double-sided silicon strip detectors to measure the $xy$ and $rz$
coordinates~\cite{coordinates} of charged particles. The three layers are 
at radii of 2.35, 3.25, and 4.75~cm. 
There is a total of 0.016 radiation lengths in the  material in the SVX
and the beryllium beam pipe whose inner radius is 1.875 cm.  For this
detector, the
average ``signal-to-noise'' ratio for charged particles at minimum 
ionization is 15:1 for the $xy$ view and 10:1 for the $rz$
view and the efficiency to have two or more SVX hits simultaneously
in both views is 95\% per track.
The impact parameter resolutions as functions of momentum $p$ (GeV/$c$) are
measured from data to be
$\sigma_{xy} = 19 \oplus 39 / (p\sin^{3/2}\theta)$\
$\mu $m and (at $\theta = 90^\circ$) 
$\sigma_{rz} = 50 \oplus 45 / p$\ $\mu $m~\cite{impact}.
The Monte Carlo simulation (MC) of the CLEO detector response is
based upon GEANT~\cite{geant}. Simulated events are processed in a
similar manner as the data.

\begin{figure}[htb]
\centering
\epsfig{figure=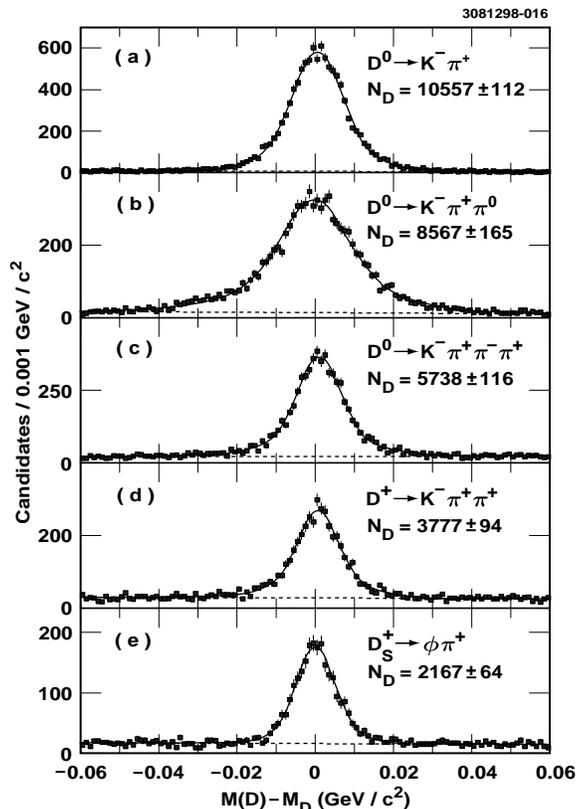,width=3.0in}
\caption{\label{fig:mass}
Masses of charmed meson candidates $M(D)$ minus the nominal masses $M_D$ for
(a) $D^0 \to K^-\pi^+$,
(b) $D^0 \to K^-\pi^+\pi^0$,
(c) $D^0 \to K^-\pi^+\pi^-\pi^+$,
(d) $D^+ \to K^-\pi^+\pi^+$,
and 
(e) $D^+_s \to \phi\pi^+$.
The data (solid squares) are overlaid with the fit to two
Gaussians with the same mean over a linear background (solid line). The
fitted background is indicated by the dashed line.} 
\end{figure}

We reconstruct $D$ mesons in the decay modes $D^0\to K^-\pi^+$,
$D^0\to K^-\pi^+\pi^0$, $D^0\to K^-\pi^+\pi^-\pi^+$,   
$D^+\to K^-\pi^+\pi^+$, and $D^+_s\to \phi\pi^+$ with 
$\phi\to K^+K^-$. 
In this
Letter, ``$D$'' refers to $D^0$, $D^+$, and $D^+_s$ mesons and reference to
the charge conjugate state is implicit. 
The charged $D$ daughters are required to 
have well reconstructed tracks and to have particle identification
information from specific ionization ($dE/dx$) and time-of-flight
consistent with the $D$ daughter hypothesis. Charged tracks forming a $D$
candidate are required to  originate from a common vertex. Neutral pions
are reconstructed from photon  pairs detected in the electromagnetic
calorimeter. The photons are required to have an energy of at least 30
(50)~MeV in the barrel (endcap) region and their
invariant  mass is required to be within three standard deviations of
the nominal $\pi^0$ mass. The $\pi^0$ momentum for 
$ D^0 \to K^-\pi^+\pi^0$ is required to be greater than  100~MeV/$c$. 
For background suppression, a soft pion $\pi_s^+$ ($\pi_s^0$) is required
to form a $D^{*+}$ with the $D$ candidate for the $D^0$ ($D^+$) decay
modes. The $D^{*+}$ -- $D^0$ ($D^+$) mass difference is required to be within
800 (1400) keV/$c^2$ of the nominal value~\cite{pdg}. No such
requirement is made for the decay $D^+_s \to\phi\pi^+$, where the requirement
that the $K^+K^-$ invariant mass be within 6~MeV/$c^2$ of the $\phi$ mass
substantially reduces the background contribution. In the
decay  $D^+_s \rightarrow\phi\pi^+$, the angle between one of the kaons and
the pion in the rest frame of the $\phi$ meson follows a
$\cos^2\theta$ distribution. Since the combinatorial background for this
decay is distributed uniformly, we require  $|\cos\theta|>0.4$. 
The $D^{*+}$ and the $D^+_s$ momenta are required to be greater than
2.5~GeV/$c$. The mass distributions for the $D$ candidates (after
subtracting the nominal $D$ mass values~\cite{pdg}) are shown in
Fig.~\ref{fig:mass}. The numbers of reconstructed $D$ mesons $N_D$,
given in the figures, are obtained from fits of the mass distributions
to two Gaussians over a linear background. The background fractions
in the mass regions within $\pm $~16~MeV of the nominal $D$ mass
values~\cite{pdg} are 
1.2\% ($D^0\rightarrow K^-\pi^+$),  
4.9\% ($D^0\rightarrow K^-\pi^+\pi^0$),  
10.0\% ($D^0\rightarrow K^-\pi^+\pi^-\pi^+$),  
12.2\% ($D^+$), and  
13.8\% ($D^+_s$).  

The dimensions of the CESR luminous region (beam spot) are
known from the machine optics to be about 1~cm along the beam direction
($z$), 7~$\mu $m in the  $y$ direction, and about 350~$\mu $m in the $x$
direction. The centroid of the beam spot is determined~\cite{beam-spot}
for each CESR fill. 
The  charm hadrons are produced approximately back-to-back
at the primary vertex (interaction point).
In the laboratory frame the selected $D^0$, $D^+$, and $D^+_s$ mesons
have an average momentum of 3.2~GeV$/c$ and average decay lengths of
200~$\mu $m, 500~$\mu $m, and 240~$\mu $m, respectively. 

\begin{figure}[htb]
\centering
\epsfig{figure=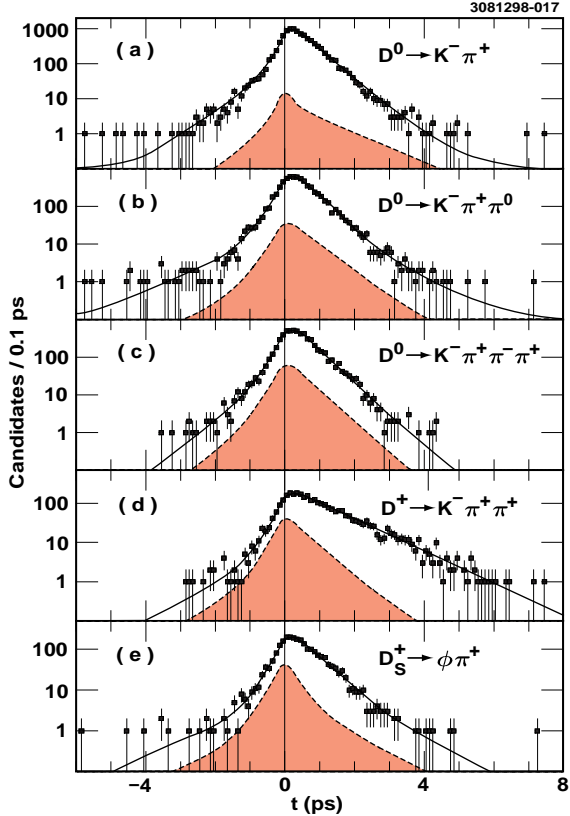,width=3.0in}
\caption{\label{fig:propt}
Proper-time distributions of charm meson candidates within $\pm 16$~MeV/$c^2$
of  the nominal $D$ mass for
(a) $ D^0 \to K^-\pi^+$,
(b) $ D^0 \to K^-\pi^+\pi^0$,
(c) $ D^0 \to K^-\pi^+\pi^-\pi^+$,
(d) $D^+ \to K^-\pi^+\pi^+$,
and 
(e) $D^+_s \to \phi\pi^+$.
The data (solid squares) are overlaid with the result from the unbinned
likelihood lifetime fit (solid line). The proper-time spectra of
the background candidates obtained from the fits are indicated by the
shaded area.}
\end{figure}

The decay vertex ${\bf r}_D$ and the momentum vector ${\bf p}_D$
of each $D$
meson candidate are reconstructed in the $xy$ plane.  The decay 
vertex resolution along the $D$ flight direction is
$80-100$~$\mu $m depending on the decay mode. The
interaction point ${\bf r}_{\rm IP}$ is reconstructed by extrapolating the 
$D$ momentum back from the decay vertex to the beam spot.  We
calculate the projected decay length $l_{\rm dec}$ from the displacement
in the $xy$ plane from ${\bf r}_{\rm IP}$ to the $D$ decay vertex,
$l_{\rm dec} = ({\bf r}_{D}-{\bf r}_{\rm IP})\cdot \hat{{\bf p}}_D$.
We then calculate the proper time of the $D$ meson decay from
$t = m_D l_{\rm dec}/c p_D$ using the appropriate PDG~\cite{pdg} world 
average for $m_D$. The proper-time distributions for the $D$ candidates 
are shown in Fig.~\ref{fig:propt}. 
 
The $D$ meson lifetimes are extracted from the proper-time distributions with
an unbinned likelihood fit. The likelihood function is
\widetext
\begin{eqnarray}\label{eq:likelihood}
  L&&(\tau_{D}, 
   f_{\rm bg}, 
  \tau_{\rm bg}, 
  S, 
  f_{\rm mis}, 
  \sigma_{\rm mis}, 
  f_{\rm wide} ) \nonumber\\
  &&=\prod_i \int_0^\infty dt^\prime 
  \left[
    \underbrace{p_{{\rm sig},i}E(t^\prime|\tau_{D})}_{\rm signal\ fraction} +
    \underbrace{(1-p_{{\rm sig},i})\left[ f_{\rm bg} E(t^\prime|\tau_{\rm bg})
    + (1- f_{\rm bg})\delta(t^\prime)\right]}_{\rm background\ fraction}
  \right] \nonumber \\
  &&\ \ \ \times \left[
    \underbrace{(1-f_{\rm mis}-f_{\rm wide})G(t_i-t^\prime|S
    \sigma_{t,i})}_{\rm proper-time\ resolution} +
       \underbrace{
         f_{\rm mis} G(t_i-t^\prime|\sigma_{\rm mis}) +
         f_{\rm wide}G(t_i-t^\prime|\sigma_{\rm wide})}_{\rm mismeasured\ fraction}
  \right]
  \nonumber
\end{eqnarray}\narrowtext

\noindent where the product is over the $D$ meson candidates, 
$G(t|\sigma)\equiv \exp(-t^2/2\sigma^2)/\sqrt{2\pi}\sigma$,
and $E(t|\tau) \equiv \exp(-t/\tau)/\tau$.
We fit the proper-time distributions for the different decay modes
separately. In these fits, each $D$ meson candidate is assigned a signal
probability $p_{{\rm sig},i}$ based on its mass. The signal
probabilities are derived from the (independent) fits of the $D$ mass
distributions to the sum of two Gaussians with the same mean and 
a linear background function. The seven parameters of the lifetime fit
are $\tau_{D}$,  $f_{\rm bg}$, $\tau_{\rm bg}$, $S$, 
$f_{\rm mis}$, $\sigma_{\rm mis}$ and $f_{\rm wide}$. 
The parameter $\tau_D$ is the $D$ meson lifetime. The background
proper-time distribution is modeled by a fraction $f_{\rm bg}$  with a
background lifetime $\tau_{\rm bg}$ and a fraction with zero
lifetime. In order to estimate the background properties, we fit
the candidates in a wide region of $\pm 40$~MeV/$c^2$ around the nominal
$D$ mass. Each candidate is weighted in the fit according to its proper-time
uncertainty $\sigma_{t,i}$. The fit allows for a global scale factor $S$
that modifies 
the calculated proper-time uncertainty.  The fits yield $S \sim 1.1$ 
for all modes.
For a small
fraction of mismeasured candidates $f_{\rm mis}$, the fitted uncertainty
$S \sigma_{t,i}$ underestimates the true uncertainty. This is a result
of track reconstruction errors such as hard multiple scattering or the
use of an SVX noise hit in the track fit. In the fit, we account for
the mismeasured candidates with two Gaussians. The fit parameters
associated with the mismeasured candidates are the fraction of events
in each of the Gaussians $f_{\rm mis}$ and $f_{\rm wide}$ and the width
of one of the 
Gaussians $\sigma_{\rm mis}$. The width of the other Gaussian
($\sigma_{\rm wide} = 8$~ps) is fixed. 
%We do not use a Monte Carlo simulation to model the proper-time distributions. 
The results of the
unbinned likelihood fits are superimposed on the proper-time distributions
shown in Fig.~\ref{fig:propt}.

From the fits we obtain 
$\tau_{ D^0} = 411.1 \pm 5.7$~fs ($K^-\pi^+$),
$395.2 \pm 8.1$~fs ($K^-\pi^+\pi^0$),
$416.3 \pm 8.6$~fs ($K^-\pi^+\pi^-\pi^+$),
$\tau_{D^+} = 1033.6 \pm 22.1$~fs,
and
$\tau_{D^+_s} = 486.3 \pm 15.0$~fs, where the uncertainties are statistical
only. 
The correlation coefficients between the $D$ lifetime and the other fit
parameters are typically near 0.1, and the largest is 0.28. 
All of these fit results have been corrected for small biases observed in
the measurements of the $D$ lifetimes in simulated events of
$-3.0 \pm 0.9$~fs ($K^-\pi^+$),
$2.4 \pm 2.3$~fs ($K^-\pi^+\pi^0$),
$-2.0 \pm 2.2$~fs ($K^-\pi^+\pi^-\pi^+$),
$-2.9 \pm 6.6$~fs ($D^+$),
and
$-0.6 \pm 2.4$~fs ($D^+_s$).
The $D^0$ lifetime $\tau_{ D^0}=408.5\pm 4.1$~fs (combined) is obtained as
the weighted average of the three measurements using statistical
uncertainties only. 
The individual measurements are weighted by their inverse relative
uncertainty $(\tau/\sigma_\tau)^2$~\cite{tau-weight}. 

The large samples of reconstructed charm mesons 
permit a number of consistency checks, including
varying the $D$ candidate mass region,
measurement of the background properties in the $D$ mass sidebands, and 
division of the data
samples in several key variables such as azimuthal angle, polar angle,
momentum of the $D$ candidate, and data taking period. No statistically 
significant
effect is found in any of these variables.

\widetext\begin{table}[htb]
  \begin{center}
  \caption{Systematic uncertainties for the $D$ meson
    lifetimes in fs.  The systematic uncertainties for the three $D^0$ 
    modes are weighted with the same weights as the fitted $D^0$
    lifetimes.}\label{tab:syst-tot-all}
    \medskip  
    \begin{tabular}{l|ccc|c|c|c}
      Uncertainty & \multicolumn{3}{c|}{$D^0$}& $D^0$ & 
                    $D^+$ & $D^+_s$ \\
      & $K^-\pi^+$ & $K^-\pi^+\pi^0$ & $K^-\pi^+\pi^-\pi^+$~~ &
      combined~~ &
      $K^-\pi^+\pi^+$~~ & $\phi\pi^+$ \\
      \hline
      Decay vertex & $\pm 2.0$  & $\pm 2.0$ & $\pm 2.0$ & 
                     $\pm 2.0$  & $\pm 2.8$ & $\pm 2.1$ \\
      Global detector scale & $\pm 0.1$ & $\pm 0.1$ & $\pm 0.1$ & 
                              $\pm 0.1$ & $\pm 0.1$ & $\pm 0.1$ \\
      Beam spot           & $^{+0.3}_{-0.1}$ & $^{+2.1}_{-0.0}$ &
                            $^{+0.3}_{-0.2}$ & $^{+0.8}_{-0.1}$ & 
                            $^{+1.3}_{-1.1}$ & $^{+0.7}_{-1.1}$ \\
      $D$ meson mass & $\pm 0.1$ & $\pm 0.1$ & $\pm 0.1$ & 
                       $\pm 0.1$ & $\pm 0.3$ & $\pm 0.1$ \\
      $D$ meson momentum & $^{+0.2}_{-0.0}$ & $^{+0.1}_{-0.2}$ &
                           $^{+0.3}_{-0.1}$ & $^{+0.2}_{-0.1}$ & 
                           $^{+0.6}_{-0.0}$ & $\pm 0.1$ \\
      Signal probability & $^{+0.4}_{-0.1}$ & $^{+0.1}_{-0.2}$ &
                           $^{+0.1}_{-0.2}$ & $^{+0.3}_{-0.1}$ &  
                           $^{+1.2}_{-8.1}$ & $^{+1.3}_{-1.8}$ \\
      $t$ -- $M(D)$ correlation & $\pm 0.6$ & $\pm 0.6$ & $\pm 1.0$ & 
                                  $\pm 0.7$ & $\pm 1.7$ & $\pm 1.5$ \\
      Large proper times & $\pm 1.2$ & $\pm 3.4$ & $\pm 0.2$ & 
                           $\pm 1.5$ & $\pm 0.3$ & $\pm 0.5$ \\
      Background & $\pm 0.5$ & $\pm 2.4$ & $\pm 3.0$ & $\pm 1.5$ & 
                   $\pm 6.3$ & $\pm 2.9$ \\
      MC statistics & $\pm 0.9$ & $\pm 2.3$ & $\pm 2.2$ & $\pm 1.6$ & 
                      $\pm 6.6$ & $\pm 2.4$ \\
      \hline
      Total &  $^{+2.7}_{-2.6}$ & $^{+5.6}_{-5.2}$ & $\pm 4.4$ &
               $^{+3.5}_{-3.4}$ & $^{+~9.9}_{-12.7}$ & $^{+4.9}_{-5.1}$ \\ 
    \end{tabular}
  \end{center}
\end{table}\narrowtext

The systematic uncertainties for the $D$ meson lifetimes are listed in
Table~\ref{tab:syst-tot-all} and are described below. They can be grouped
into three categories: 

{\it Reconstruction of the $D$ decay length and proper time.}
Errors in the measurement of the reconstructed decay length can be due to
errors in the measurement of the decay vertex,  the global detector scale,
and the beam spot.   The bias in the decay vertex position is
estimated to be ($0.0 \pm 0.9~\mu $m)
from a ``zero-lifetime'' sample of 
$\gamma\gamma \to \pi^+\pi^-\pi^+\pi^-$ events. This  corresponds to a
measured proper-time uncertainty of $\pm 1.8$~fs. In addition, the vertex
reconstruction is checked with events with interactions in the beam pipe with
a relative uncertainty of $\pm 0.2$\%. The sums of these uncertainties in
quadrature  yield the systematic uncertainties due to the decay vertex
measurement.   
The global detector scale is measured to a precision of $\pm 0.1$\% in surveys
and confirmed in the study of events with interactions in the beam pipe.
The changes in the lifetimes due to the variation ($\pm $2~$\mu $m) in the
vertical beam spot position and height are another source of systematic
error, since the interaction point is calculated from the beam spot and the
reconstructed $D$ momentum and decay vertex.
Statistical uncertainties for the $D$ masses~\cite{pdg} and the $D$ momentum
measurements lead to systematic errors since these quantities are used
to convert the decay length into proper time.  

{\it Lifetime fit procedure.}
This category includes uncertainties in the candidate signal
probabilities, the impact of candidates with large proper times, the 
correlation between proper time and $D$ meson mass, and the proper-time
properties of the background. 
The signal probability assigned to each candidate in the
lifetime fit has a statistical uncertainty, and these statistical
uncertainties lead to systematic uncertainties in the fitted lifetimes. 
We estimate these systematic uncertainties by coherently varying the signal
probability of each candidate by its statistical uncertainty and repeating
the fits.   
A correlation between the measurements of the proper time $t$ and the
$D$  candidate mass $M(D)$
can be a source of systematic uncertainty.  
We measure this correlation in simulated events to estimate the
associated systematic uncertainty.
Charm meson candidates with large proper times are an additional source of 
systematic uncertainty. These candidates are modeled by the
wide Gaussian in the proper-time fit.  Alternatively, the wide Gaussian
component is omitted from the likelihood function and candidates 
in a restricted proper-time interval are fitted. 
The systematic uncertainties due to
candidates with large proper times are estimated from the variations of
$\tau_D$ with the width of the wide Gaussian and the differences in the
results between the fits with different proper-time intervals.   
This systematic
uncertainty is small for decay modes with three or more charged
$D$ daughters for which the requirement of a well-reconstructed vertex
greatly reduces mismeasurements.
We estimate the systematic uncertainty due to charm and other backgrounds
that might populate the $D$ mass peaks differently than they
populate the $D$ mass sidebands,
$20~{\rm MeV}/c^2 < |M(D) - M_D| < 60~{\rm MeV}/c^2$.
Some possible sources of such backgrounds are a background in the $D_s^+$
sample  from $D^+ \to K^+\pi^-\pi^-$ decays where one $\pi^-$ is
misidentified as a $K^-$, and backgrounds from $D^{+(0)}$ decays in the
$D^{0(+)}$ sample caused by adding or missing a charged pion.

{\it Checking the algorithms with simulated events.}
Charm meson candidate selection requirements can cause systematic biases in
the lifetime measurements.  We estimate these biases with simulated events
and correct for the biases as described above. We include the statistical
uncertainties in the measured lifetimes from the samples of simulated events
as systematic uncertainties in the results.  

The total systematic uncertainties in the $D^0$ lifetime measurement
are obtained by combining the contributions from the three reconstructed
$D^0$ decay modes. The contributions from the decay length measurement
and the detector size are assumed to be completely correlated and all
other contributions are assumed to be uncorrelated. The total
systematic uncertainties are obtained by adding the individual
contributions in quadrature. 

In summary, we report measurements of charm meson lifetimes
from 3.7~fb$^{-1}$ of integrated luminosity recorded with 
the CLEO detector. The measured $D$ lifetimes are
$\tau_{D^0} = 408.5 \pm 4.1 ^{+3.5}_{-3.4}$~fs, 
$\tau_{D^+} = 1033.6 \pm 22.1 ^{+~9.9}_{-12.7}$~fs, and 
$\tau_{D^+_s} = 486.3 \pm 15.0 ^{+4.9}_{-5.1}$~fs,
where the first uncertainties are statistical and the
second systematic. 
These results imply $\tau_{D^+_s}/\tau_{D^0} = 1.19 \pm 0.04$, a difference
of more than 4.5 standard deviations in a single experiment.
The charm meson lifetimes reported in this Letter
are comparable in precision with the best previous
measurements~\cite{fixed-target}, and the systematic
errors are very different.  

We gratefully acknowledge the effort of the CESR staff in providing us with
excellent luminosity and running conditions.
We wish to acknowledge and thank the technical staff who contributed to
the success of the CLEO~II.V detector upgrade, including J.~Cherwinka and
J.~Dobbins (Cornell); M.~O'Neill (CRPP); M.~Haney (Illinois); M.~Studer and 
B.~Wells (OSU); K.~Arndt, D.~Hale, and S.~Kyre (UCSB).
We appreciate contributions from G.~Lutz and advice from A.~Schwarz.
This work was supported by 
the National Science Foundation, 
the U.S. Department of Energy,
Research Corporation,
the Natural Sciences and Engineering Research Council of Canada, 
the A.P. Sloan Foundation, the Swiss National Science Foundation, 
and the Alexander von Humboldt Stiftung.


\begin{thebibliography}{99}

\bibitem{theory}
G.~Bellini, I.~Bigi and P.J.~Dornan, Phys. Rept. {\bf 289}, 1 (1997).
\bibitem{pdg} 
Particle Data Group, C.~Caso {\em et al.}, Eur.~Phys.~J.~C~{\bf 3}, (1998). 
\relax
\bibitem{fixed-target} 
E687 Collaboration,  P.L.~Frabetti {\em et al.}, Phys. Rev. Lett.~{\bf
  71}, 827~(1993).\\ % Ds
E687 Collaboration,  P.L.~Frabetti {\em et al.}, Phys. Lett.~B {\bf
  323}, 459~(1994).\\ % D0, D+
E691 Collaboration, J.R.~Raab {\em et al.}, Phys. Rev. D~{\bf 37},
2391~(1988).\\  %  D0, D+, Ds \relax
E791 Collaboration, E.M. Aitala {\em et al.}, Phys. Lett.~B {\bf 445},
449~(1999). % Ds
\bibitem{cleoii} 
CLEO Collaboration, Y. Kubota {\em et al.}, Nucl. Instrum. Methods A {\bf
  320}, 66 (1992). \relax
\bibitem{svx}
T. Hill, Nucl. Instrum. Methods A {\bf 418}, 32 (1998). \relax
\bibitem{coordinates}
The right handed coordinate system has the $z$ axis along the $e^+$ beam
direction and the $y$ axis upward.
\bibitem{impact}
Later improvement of the track-fitting code resulted in an $rz$ impact
parameter resolution of $\sigma_{rz} = 42 \oplus 45 / p$\ $\mu $m at
$\theta = 90^\circ$. 
\bibitem{geant}
  R. Brun {\em et al.}, GEANT 3.15, CERN Report No.~DD/EE/84-1 (1987). \relax
\bibitem{beam-spot}         
  D.~Cinabro {\em et al.}, Phys. Rev. E {\bf 57}, 1193 (1998). \relax
%\bibitem{tau-weight}         
%  L.~Lyons, A.J.~Martin and D.H.~Saxon, Phys. Rev. D {\bf 41}, 982 (1990). \relax
\bibitem{tau-weight}         
  L.~Lyons and D.H.~Saxon, Rep. Prog. Phys. {\bf 52}, 1015 (1989). \relax
\end{thebibliography}
\end{document}